\documentclass[12pt,english]{article}
\pdfoutput=1
\usepackage{graphicx,array}
\usepackage[margin=0.9in]{geometry}
\usepackage{cite}
\usepackage{color}
\usepackage{latexsym}
\usepackage{amsthm}
\usepackage{amsmath}
\usepackage[titletoc]{appendix}
\usepackage{enumitem}
\usepackage{amssymb}
\usepackage{wrapfig}

\usepackage[unicode=true,
 bookmarks=true,bookmarksnumbered=false,bookmarksopen=false,
 breaklinks=false,pdfborder={0 0 1},backref=false,colorlinks=true]
 {hyperref}
\hypersetup{pdftitle={Probing the Quantum Nature of Gravity in the Microgravity of Space},
 pdfauthor={Singh},
 citecolor=blue,linkcolor=blue,urlcolor=blue,citecolor=blue,linkcolor=blue}
\usepackage{breakurl}

\def\simgt{\mathrel{\lower2.5pt\vbox{\lineskip=0pt\baselineskip=0pt
           \hbox{$>$}\hbox{$\sim$}}}}
\def\simlt{\mathrel{\lower2.5pt\vbox{\lineskip=0pt\baselineskip=0pt
           \hbox{$<$}\hbox{$\sim$}}}}

\newcommand{\be}{\begin{equation}}
\newcommand{\ee}{\end{equation}}
\newcommand{\bea}{\begin{eqnarray}}
\newcommand{\eea}{\end{eqnarray}}

\definecolor{nicered}{rgb}{0.7,0.1,0.1}
\definecolor{nicegreen}{rgb}{0.1,0.5,0.1}
\hypersetup{colorlinks,citecolor=black,linkcolor=black,urlcolor=black}

\definecolor{purple}{rgb}{0.5,0,0.5}
\definecolor{burgundy}{rgb}{0.5, 0.00, 0.13}

\usepackage{lipsum}

\newcommand\blfootnote[1]{%
  \begingroup
  \renewcommand\thefootnote{}\footnote{#1}%
  \addtocounter{footnote}{-1}%
  \endgroup
}

\begin{document}
\baselineskip=14pt
\hfill

\vspace{2cm}
\thispagestyle{empty}
\begin{center}
{\Large\bf
Topical: Probing the Quantum Nature of Gravity in the Microgravity of Space
}\\
\bigskip\vspace{0.5cm}{
{\large Ashmeet Singh}\blfootnote{White Paper written for the \href{https://www.nationalacademies.org/our-work/decadal-survey-on-life-and-physical-sciences-research-in-space-2023-2032}{National Academies' Decadal Survey on Biological and Physical Sciences (BPS) Research in Space 2023-2032}  \\}
} \\[7mm]
 {\it 
Department of Physics\\
    California Institute of Technology \\
    1200 E. California Blvd., Pasadena, CA 91125 USA\\
    Email: \href{mailto:ashmeet@caltech.edu}{ashmeet@caltech.edu}\\
    Phone: +1-626-787-4819\\~\\	
    }

 \end{center}
\bigskip
\vspace{-0.5cm}
\centerline{\large\bf Abstract}

\begin{quote} \small
 Conventionally, experiments probing the quantum nature of gravity were thought to be prohibitive due to the extremely high energy scales involved. However, recent and rapid advances at the intersection of quantum information and gravity, along with quantum technologies that allow preparation and control of mechanical systems in the quantum regime, indicate that such tests may well be within reach of upcoming experimental capabilities. The microgravity of space offers a unique environment to carry out this endeavor, allowing possibilities to control and manipulate delicate quantum characteristics in larger systems, better than current Earth-based setups. In this white paper, we lay out the science case for furthering a community effort to study and lead progress in both theoretical and experimental aspects for space-based tests of fundamental physics, particularly to probe the elusive quantum nature of gravity.
\end{quote}

\vspace{1.5cm}

\newpage
\baselineskip=16pt
	
\setcounter{footnote}{0}
\setcounter{page}{1}

\section{Motivation} \label{sec:Motivation}
Quantum mechanics and gravity have emerged as the two key pillars of our fundamental understanding of the universe, and have been tremendously successful in making quantitative predictions for experiments in their respective domains. 
Quantum physics forms the backbone of the strong force, weak force, and electromagnetic force, however, quantum mechanics and gravity find themselves at strong odds in their current formulation, both conceptually and mathematically. Finding a theory of gravity consistent with quantum principles is one of the most defining and important problems in modern physics. The questions we face are profound and foundational, touching upon the quantum origins of space and time.

\vspace{0.1cm}
Conventionally, experiments to study the quantum nature of gravity have been considered prohibitive due to the extremely high energy, or correspondingly, extremely tiny length scales involved. It is traditionally expected that quantum gravitational effects would become relevant around the Planck length, $l_{\mathrm{pl}} = \sqrt{\hbar G/c^3} = 1.62 \times 10^{-35} \: \mathrm{m}$. This is indicative of the ``extremes'' involved: quantum effects (represented by the reduced Planck's constant $\hbar$) are most pronounced at the microscopic scale, and gravity (represented by $G$, Newton's gravitational constant) being the weakest of all the fundamental forces which becomes important at the macroscopic scale ($c$ is the speed of light in vacuum). At the Planck scale, one expects to witness \emph{quantum effects in spacetime itself}. However, to reach such high energies using traditional collider physics, one would need to build a Milky Way-sized particle accelerator\cite{Hossenfelder2018}. Astrophysics and cosmology\cite{Barrau:2017tcd,PhysRevD.93.023505} offer some avenues to probe quantum gravity, such as early universe physics, or neutron stars and black holes, but they are often obscured by the lack of laboratory-like control and the presence of various interfering, competing effects.

\vspace{0.1cm}
We thus require a paradigm where we can study quantum effects in low energy, meso-to-macroscopic mechanical systems, and by extension, their gravitational interactions, in a controlled setting. Quantum information theoretic techniques and quantum sensing using ultracold atoms, light-matter interaction, atom interferometry, optomechanical resonators, among others, allows us to do just that. Instead of looking for quantum effects in spacetime at the Planck length, we let gravity act as a mediator which can dynamically and decisively process quantum information, such as looking for signatures of quantum entanglement, decoherence, etc. Recent breakthroughs in theoretical quantum information and complementary advances in preparation, control, and measurements of quantum systems suggest that such experiments may very well be possible in the near future. 

\vspace{0.1cm}
Quantum properties and their signatures are fragile. The microgravity of space offers a unique environment to carry out this endeavor, allowing possibilities to control and manipulate quantum properties in larger systems, better than current Earth-based setups. In this white paper, we lay out the science case for furthering a community effort to study both theoretical and experimental aspects of quantum information and sensing in space-based tests of fundamental physics, particularly the elusive quantum nature of gravity. 
\vspace{-0.3cm}
\section{Science Case} \label{sec:science}
\vspace{-0.2cm}
\subsection{Theoretical Background and Setup}
Quantum mechanics is not just a different theory of how fundamental particles and forces operate, but rather a different \emph{kind} of theory, representing a stark departure from the classical, Newtonian paradigm. The uniqueness of quantum physics broadly lies in the following properties:
\begin{itemize}[topsep=0pt,itemsep=-.05ex,partopsep=1ex,parsep=1ex]
    \item \textbf{Superposition:} Quantum systems need not be in particular, definitive classical configurations, rather they can exist in admixtures of various classical possibilities.
    \item \textbf{Quantum Entanglement:} is the presence of correlations in multi-particle quantum systems such that the state of any one particle cannot be completely described independent of the state of the others. 
    \item \textbf{Non-commutativity:} implies the existence of incompatible observables (those for which the system does not have simultaneously existing values), and the order of operations on a quantum system matters. In general, $A B \neq B A$.
    \item \textbf{Interactions and Measurements:} Quantum states evolve linearly by the Schrödinger's equation, a first order differential equation in time. Observations/measurements effectively alter the quantum state to produce a conclusive outcome, the process of which is not fully understood, however.
\end{itemize}
\vspace{0.2cm}
On the other hand, our best understanding of gravity comes from Einstein's general theory of relativity, a classical field theory, obeying the principles of Newtonian determinism. It has been extremely robust to experimental tests, from millimeter\cite{Westphal:2020okx} scales all the way to cosmological ones\cite{WMAP:2003elm}. It remains unclear whether gravity is fundamentally a quantum entity or a classical one\cite{Carney:2018ofe} (and references therein). Furthermore, it remains unclear if either quantum mechanics and/or gravity need to alter their fundamental principles. Theoretical progress alone seems unlikely to settle this question, and hence we need to turn towards empirical support and guidance. To understand quantum aspects of gravity, or the lack thereof, we must test whether gravity has one or more of the defining principles of quantum theory listed above. Whether gravity is communicated or mediated via a quantum channel or a classical one, or something altogether different, can shed light on the quantum nature of gravity.

A keystone question to investigate is whether a mass in a superposition of two locations produces a gravitational field also existing in a superposition? One way to study this question is to detect whether gravity can lead to entanglement generation between two quantum systems\cite{Diosi:1984wuz,Kafri:2014hlh,Bose:2017nin,Marletto:2017kzi,Haine:2018bwu,Chevalier:2020uvv,Anastopoulos:2020cdp,Matsumura:2020law}, which would nominally only be possible if it were a quantum interaction. A few such promising experimental tests, often known as ``entanglement witnesses'' \cite{Horodecki:1996nc,TERHAL2000319} (of which a particular example is the test of violation of the famous Bell's inequalities) have been proposed in the literature. To verify entanglement generation, these setups often require measurements on both quantum systems involved, which makes implementation challenging. Other recent proposals suggest studying gravitationally generated entanglement between quantum systems near measurement events\cite{Kent:2020gov}. On the other hand, lack of entanglement production would hint towards classical models of gravity, which are often based on non-linear mechanisms of feedback and measurements, and can potentially leave imprints in experiment too\cite{helou2017measurable,kafri2014classical,helou2017}. 

One can also study the decoherence induced in a superposed quantum system due to gravitational interaction. As far as we understand, gravity is a universal force and cannot be completely shielded, and this can lead to loss of coherence of an initially highly coherent quantum state. Depending on whether gravity operates quantum mechanically or classically, phenomenologically distinct outcomes in terms of entropy generation, dephasing\cite{Kafri:2014hlh}, periodic revival of coherence\cite{Carney:2021yfw}, heating, time scales involved, etc., can be used to constrain models of gravitational interaction\cite{kafri2014classical,Bassi:2017szd,Anastopoulos:2013zya,Oniga:2015lro,Tilloy:2017ael}. An attractive feature of many such experiments is that they can be performed on single quantum systems.

Another promising set of ideas hinge on letting gravity carry out tasks only a quantum computer could. For example, if gravity were a quantum entity, it would process quantum information in a way which could generate non-Gaussianity in an otherwise gaussian initial state, something a classical interaction could not do, and this can be probed by studying its statistics of the quantum state\cite{Howl:2020isj}. 

Gravity can also possibly alter the fundamental canonical commutation relations between position and momentum in quantum mechanics, and can lead to modifications to Heisenberg's uncertainty principle \cite{Maggiore:1993rv,Bosso:2017hoq}, which can be detected in experiment.
\vspace{-0.2cm}
\subsection{Experimental Setups}
\begin{figure}[h!]
\centering
  \includegraphics[width=0.9\textwidth]{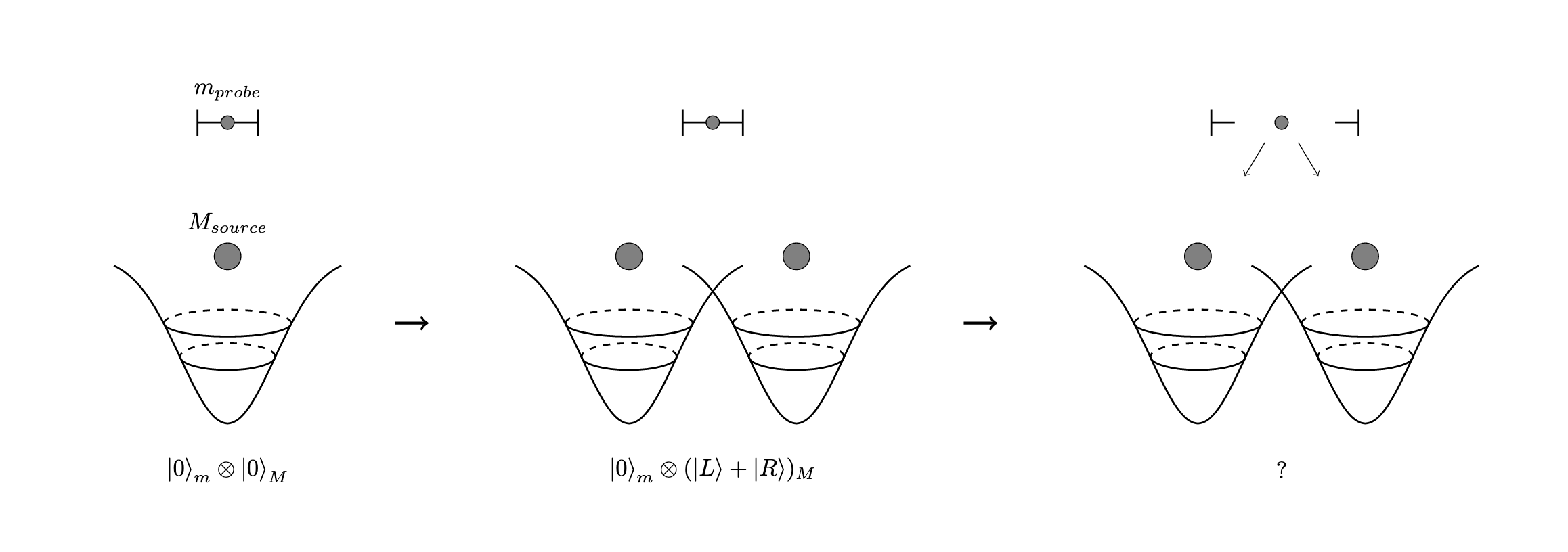}
   \caption{Schematic representation of the experimental paradigm to probe the quantum nature of gravity as explained in the text. Figure reused with permission from the authors of Carney et al. \cite{Carney:2018ofe}. }
  \label{fig:basic_setup}
\end{figure}

The basic experimental setup needed, schematically shown in Figure \ref{fig:basic_setup}, involves the preparation of a ``source'' matter in a highly quantum state -- such as being in a superposition of two locations, or being in a non-classical state (for example, the coherent states of a mechanical oscillator) -- and then study the gravitational effect it has on a ``probe'' system (which could be the source itself too).

Broadly, two possible experimental setups have been proposed: (i.) interferometric tests using quantum matter, and (ii.) non-interferometric ones, particularly optomechanical setups. An important difference between these two setups is the timescales involved: in interferometric setups, the time the particle takes to traverse the interferometer's arms sets the scale, whereas for optomechanical ones, it is the oscillator's period of oscillation.
\subsubsection{Interferometric Tests}
Interferometric techniques\cite{Geiger:2020aeq} offer a natural paradigm to study coherent quantum systems in superposition. A massive particle is put in a quantum superposition, typically separating the wave packet to exist in two locations, by passing it through a beam-splitter. The two locations of the superposition act as two arms of an interferometric apparatus. The particle is then allowed to propagate for some time, before it is recombined using another beam-splitter, and is measured to yield statistical characteristics of the evolved quantum state. By studying the interference pattern between the two components of the superposition, it can allow us to detect gravitationally induced entanglement and decoherence due to interactions with other systems. Based on differential gravitational environments in the two arms (such as a test mass closer to one arm than the other), we can also study how gravitational interaction couples with the mass in superposition. 
\\
The enabling technology is the use of atom interferometry (AI), a new weak force measurement technique based on laser cooled ultra-cold atoms. The development of AI is very active worldwide. Progress include large matter wave-packet separation of single atoms in a 10 m atomic fountain\cite{Kovachy2015QuantumSA}, and super-cold atomic samples\cite{PhysRevLett.114.143004}, among others.
\subsubsection{Non-Interferometric Tests}
A complementary setup is to use a mechanical oscillator prepared in a highly non-classical state, and use it as a witness to process gravitational interactions with a test mass. The restoring force for the oscillator  could come from a variety of possible confinement mechanisms, chief among these are optomechanical setups\cite{Chen:2013sgs,RevModPhys.86.1391,Pikovski:2011zk}, where mechanical systems are coupled with light. Other examples include optically trapped lattices, mechanical cantilevers, and high-tension membranes.
\\
The enabling technology is ultracold atomic physics which allows precise preparation and control of atoms in non-classical states, in particular using Bose Einstein condensates (BECs)\cite{1999cond.mat..4034K}. The reason one needs to cool atoms to ultracold temperatures is to prepare the particles close to their ground state configurations (which are highly non-classical, such as being extended spatially superposed wavepackets), and protect them from thermal noise.
\vspace{-0.3cm}

\section{Leveraging the Microgravity of Space} \label{sec:space}
The microgravity of space makes for a unique environment to conduct such interferometric and optomechanical experiments, allowing points of leverage that are otherwise extremely difficult to implement using terrestrial setups. The primary considerations are twofold: (i.) that quantum effects, such as coherence, superposition, entanglement, are very delicate and prone to suppression due to interactions, especially as one goes into the meso-to-macroscopic regime, and (ii.) gravity being an extremely weak force on smaller scales and tends to get dominated by other background and competing effects. On Earth, control of quantum systems and their gravitational interaction is limited by strong local background effects (including Earth's gravity) and various sources of environmental noise. In space, on the other hand, one gains leverage, for instance,
\begin{itemize}[topsep=-1pt,itemsep=-.05ex,partopsep=1ex,parsep=1ex]
    \item Less environmental decoherence allows longer coherence times to preserve fragile quantum superpositions to conduct experiments on.
    \item Suppression of Earth's gravity, along with lack of low frequency Newtonian (or gravity gradient) noise which arises from local fluctuations in the Earth's gravitational field, allows for better control and detection of gravitationally induced effects due to other controlled test masses.
    \item There is less environmental noise in space, where one has to primarily deal with the cosmic microwave background and solar radiation only.
    \item Microgravity assists in suspension of optomechanical systems, isolation from mechanical vibrations (particularly low frequency $\sim 1 - 10 \: \mathrm{Hz}$ vibrational modes) and noise, allowing for better control of the system and its systematics.
    \item Particularly for interferometric tests, one needs longer free fall times for gravitational effects to accumulate, eg. phase differences between the two arms of the interferometer. One can obtain $\sim 100 \: \mathrm{s}$ of freefall time in space compared to a few seconds on Earth.
    \item To prepare quantum systems at ultracold temperatures, the low ambient temperature of space ($\lesssim 20 \: \mathrm{K}$) is helpful and suppresses unwanted thermal noise. Additionally, the vacuum of space and low pressure conditions ($\sim 10^{-13} \: \mathrm{Pa}$) allow better control and manipulation.
    \item Better control on individual atoms allows for longer interrogation times, and hence better data statistics.
\end{itemize}
\vspace{-0.2cm}
\section{Outlook, and the Road Ahead} \label{sec:conclusion}
\vspace{-0.2cm}
What once seemed out of reach, now armed with tools of quantum information and sensing, probing the quantum nature of gravity may well be within reach of upcoming experimental probes. The microgravity of space will be a crucial platform in taking this endeavor to fruition. An strong initial thrust in this direction has already been set in action. Efforts by international space agencies towards space applications include the 100 m drop tower microgravity experiments\cite{Rudolph:2015mza}, the sounding rocket experiments by DLR\cite{2018Nature}, zero-g parabolic flights and space gradiometer project by CNES\cite{2016NatCo...713786B}, atomic interferometric gravitational wave space observatory by US\cite{Hogan:2011tsw} and Chinese space agency\cite{Gao:2017rgh}, respectively. NASA has also launched the Cold Atom Lab (CAL)\cite{jpl_cal} developed and operated by JPL, which will be the first cold atom research facility in orbit. The European Space Agency (ESA) has called for a dedicated, medium-sized space mission, such as the MAQRO proposal\cite{Kaltenbaek:2015kha,Gasbarri:2021sdm} or the STE-QUEST\cite{Altschul:2014lua}, to be conducted within the framework of the "Cosmic Vision Programme" to carry out tests of quantum physics in space.

 While the prospect of shedding light on one of the longest held, most fundamental questions in modern physics is immensely exciting, a lot of work lies ahead of us to realize this vision in space. We require a strong synergy between theoretical and experimental progress. On the theory side, we need to better model and understand the interplay between quantum information and gravity. Experimentally, we need to gain better control over preparation and manipulation of larger masses in highly quantum states, all while preparing the technology needed to implement it in space. In addition to being a ripe setting to study the quantum nature of gravity, quantum information-motivated experiments carried out in space will also allow us to study other important topics in fundamental physics, such as verification of the equivalence principle\cite{Williams:2015ima}, tests of quantum physics\cite{Gasbarri:2021sdm}, dark matter\cite{AEDGE:2019nxb}, dark energy\cite{Tino:2019tkb,2015Sci...349..849H}, and gravitational waves\cite{Hogan:2011tsw}.
\\
\\
We thus call for a collective community effort in the upcoming decade to bring together scientists and technologists, along with industry and space agencies, to outline a concrete road-map to drive and lead the effort to further our understanding of the connections between the very fabric of spacetime and the underlying quantum laws. 

\bibliographystyle{utphys}
\bibliography{ref}

\end{document}